# Quantum Logic using Linear Optics

J.D. Franson, B.C. Jacobs, and T.B. Pittman

Johns Hopkins University

Applied Physics Laboratory

Laurel, MD 20723

Abstract:

In order for quantum communications systems to become widely used, it will probably be necessary to develop quantum repeaters that can extend the range of quantum key distribution systems and correct for errors in the transmission of quantum information. Quantum logic gates based on linear optical techniques appear to be a promising approach for the development of quantum repeaters, and they may have applications in quantum computing as well. Here we describe the basic principles of logic gates based on linear optics, along with the results from several experimental demonstrations of devices of this kind. A prototype source of single photons and a quantum memory device for photons are also discussed. These devices can be combined with a four-qubit encoding to implement a quantum repeater.

## I. Introduction

Systems for quantum key distribution have been demonstrated over limited distances, both in optical fibers and in free space, but they have not yet been used for practical applications. In order for quantum communications systems to become widely used, it will probably be necessary to develop quantum repeaters that can extend the range of quantum key distribution systems and correct for errors in the transmission of quantum information. One of the most promising approaches for the development of quantum repeaters is the use of linear optical techniques [1, 2] to implement quantum logic gates, combined with optical storage loops to implement a quantum memory device for single photons [3-5]. In this chapter, we describe several prototype quantum logic gates [6-8], a single-photon source [9], and a single-photon memory device [3, 4] that we have recently demonstrated. A four-qubit encoding [5] that allows these devices to be combined to implement a quantum repeater will also be described.

The past development of quantum key distribution has been strongly influenced by the need to overcome a variety of practical challenges, and the future development of the field will probably be determined by the challenges that remain. As a result, we begin in Section II with a brief review of the challenges facing the development of quantum communications systems, both past and future. In Section III, we describe the basic operation of probabilistic quantum logic gates based on linear optics techniques, along with experimental results from several devices of that kind. The development of quantum repeaters will also require a source of single photons and a quantum memory device, and demonstrations of prototype devices of that kind are described in Section IV.

A proposed implementation [5] of a quantum repeater using a combination of these devices is outlined in Section V, followed by a summary in Section VI.

## II. Challenges in quantum communications

Quantum key distribution systems have evolved over the past 15 years in response to a number of technical challenges that limited their performance at the time. As a result, it may be useful to briefly review the past development of quantum key distribution systems and to discuss the remaining challenges that seem likely to determine the future development of the field of quantum communications.

At one time, the only known method for quantum key distribution was based on the use of the polarization states of single photons. In addition to introducing the BB84 and B92 protocols, Bennett et al. also performed the first experimental demonstration of quantum key distribution using photon polarization states in a table-top experiment [10]. But the use of single-photon polarizations was considered to be a major obstacle to practical applications at the time, since the state of polarization of a photon will change in a time-dependent way as it propagates through an optical fiber. In response to this problem, we developed a feed-back loop [11, 12] that automatically compensated for the change in polarization of the photons. The system alternated between high-intensity bursts that determined the necessary corrections to the polarization, and single-photon transmissions used for the generation of secret key material. The corrections themselves were applied using a set of Pockels cells that also controlled the transmitted polarization state in a BB84 implementation. A system of this kind [13] implemented error correction and privacy amplification in 1994, and it was the first fully-automatic and continuously operating quantum key distribution system.

Quantum key distribution systems based on an interferometric approach are now widely used. They have the advantage of being relatively insensitive to changes in the state of polarization in optical fibers. The evolution of interferometer systems of this kind is illustrated in Figure 1. The two-photon interferometer shown in Figure 1a was proposed by one of the authors in 1989 [14, 15]. Roughly speaking, two entangled photons propagate toward two distant interferometers that both contain a long path L and a short path S. The photons are emitted at the same time in a parametric down-conversion source, and if they arrive at the detectors at the same time, it follows that they both must have traveled the longer path or they both must have traveled the shorter path. Quantum interference between these two probability amplitudes gives rise to nonlocal quantum correlations that violate Bell's inequality.

As early as 1989, John Rarity noted that a two-photon interferometer of this kind could be used as a method of quantum key distribution [16]. Ekert, Rarity, Tapster, and Palma later [17] showed that tests of Bell's inequality could be used to ensure that an eavesdropper cannot determine the polarization states of the photons without being detected, which allows secure communications to be performed. Systems of this kind have now been experimentally demonstrated [18]. One potential advantage of an entangled-photon approach of this kind is that no active devices are required in order to choose a set of random bases for the measurement process. Instead, 50-50 beam splitters can randomly direct each photon toward one of two interferometers with fixed phase shifts.

The interferometric approach of Fig. 1a has the disadvantage of requiring a parametric down-conversion source, which typically has a limited photon generation rate.

Charles Bennett realized [19], however, that the need for an entangled source could be eliminated by passing a single photon through two interferometers in series, as illustrated in Fig. 1b. Although nonlocal correlations cannot be obtained in such an arrangement, it does allow the use of weak coherent state pulses containing much less than one photon per pulse on the average. The ease in generating weak coherent state pulses combined with the relative lack of sensitivity to polarization changes made this type of interferometer system relatively easy to use. As a result, a number of groups [20-23] demonstrated quantum key distribution systems of this kind, including work by Townsend, Rarity, Tapster, and Hughes.

One of the disadvantages of the interferometric approaches of Figures 1a and 1b is that the relative phase of the two interferometers must be carefully stabilized. In addition, the polarization of the photons must still be controlled to some extent in order to achieve a stable interference pattern. Gisin and his colleagues [24] avoided both of these difficulties by using a very clever technique illustrated in Fig. 1c. Here the system is essentially folded in half by placing a mirror at one end of the optical fiber and reflecting the photons back through the same interferometer a second time. By using a Faraday mirror, the state of polarization is changed to the orthogonal state during the second pass through the optical fiber, which eliminates any polarization-changing effects in the optical fiber. Plug-and-play systems of this kind are very stable and are now in widespread use.

The remaining problem in existing quantum key distribution systems is the limited range that can be achieved in optical fibers due to photon loss. As a potential solution to this problem, we performed the first demonstration [25] of a free-space system over a relatively short distance outdoors in broad daylight in 1996. The accidental detection rate due to the solar background was minimized using a combination of narrow-band filters, short time windows, and a small solid angle over which the signal was accepted. A number of other groups [26-27] have now demonstrated similar systems over larger ranges, and satellite systems of this kind are being considered. These systems will probably have relatively high costs and small bandwidths.

The widespread use of quantum communications systems will require both large bandwidth and operation over large distances. Although earlier limitations due to polarization changes in fibers and the stability of interferometric implementations have now been overcome, it seems likely that quantum repeaters [5, 28, 29] will be required in order to achieve the necessary bandwidth and operational range. A promising approach for the implementation of a quantum repeater is described in the following sections.

### III. Linear optics quantum logic gates

Quantum logic operations are inherently nonlinear, since one qubit must control the state of another qubit. In the case of photonic logic gates, this would seem to require nonlinear optical effects, which are usually significant only for high-intensity beams of light in nonlinear materials. As shown by Knill, Laflamme, and Milburn (KLM), however, probabilistic quantum logic operations can be performed using linear optical elements, additional photons (ancilla), and post-selection based on the results of measurements made on the ancilla [1].

The basic idea of linear optical logic gates is illustrated in Figure 2. Here two qubits in the form of single photons form the input to the device and two qubits emerge, having undergone the desired logical operation. In addition, a number of ancilla photons also enter the device, where they are combined with the two input qubits using linear optical elements, such as beam splitters and phase shifters. The quantum states of the ancilla are measured when they leave the device, and there are three possible outcomes: (a) When certain outcomes are obtained, the logic operation is known to have been correctly implemented and the output of the device is accepted without change. (b) When other measurement outcomes are obtained, the output of the device is incorrect, but it can be corrected in a known way using a real-time correction known as feed-forward control, which we have recently demonstrated [30]. (c) For the remaining measurement outcomes, the output is known to be incorrect and cannot be corrected using feed-forward control. The latter events are rejected and are referred to as failure events. The probability of such a failure can scale as $1/n$ or $1/n^2$, depending on the approach that is used [1, 2].

The original approach suggested by KLM was based on the use of nested interferometers [1]. It was subsequently shown [6, 31] that similar devices could be implemented using a polarization encoding, which had the advantage of simplicity and lack of sensitivity to phase drifts. A controlled-NOT (CNOT) quantum logic gate implemented in this way [6] is shown in Figure 3. Its implementation requires only two polarizing beam splitters, two polarization-sensitive detectors, and a pair of entangled ancilla used as a resource. The correct logical output is obtained whenever each detector registers one and only one photon, which occurs with a probability of ¼.

The CNOT gate shown in Figure 3 can be understood as the combination of several more elementary gates, including the quantum parity check [6, 32] shown in Figure 4. The intended purpose of this device is to compare the values of the two input qubits without measuring either of them. If the values are the same, then that value is transferred to the output of the device. If the two values are different, then the device indicates that the two bits were different and no output is produced. A quantum parity check of this kind can be implemented using only a single polarizing beam splitter and a single polarization-sensitive detector.

An experimental apparatus [7] used to implement a quantum parity check is outlined in Figure 5. Parametric down-conversion was used to generate a pair of photons at the same wavelength. In type-II down-conversion, the two photons have orthogonal polarizations, so that a polarizing beam splitter could be used to separate the photons along two different paths. Waveplates could be used to rotate the plane of polarization of the photons, which created a quantum superposition of logical states, where a horizontally-polarized photon represented a value of "0" and a vertically-polarized photon represented a value of "1". The parity check itself was implemented with a second polarizing beam splitter, after which the state of polarization could be measured using polarization analyzers and single-photon detectors. The results of the experiment [7] are shown in Figure 6 for the case in which the input qubits had definite values of 0 or 1. Here the correct results are shown in blue while incorrect results are shown in red. Similar performance was also obtained using superposition states as inputs, which demonstrates the quantum-mechanical coherence of the operation.

Another useful quantum logic gate is the quantum encoder [6] shown in Figure 7. The intended function of this device is to copy the value of a single input qubit onto two output qubits. Once again, this operation has to be performed without measuring the value of the qubits. Our implementation of a quantum encoder requires a pair of entangled ancilla photons in addition to a polarizing beam splitter. The results from an experimental demonstration [33] of a quantum encoder are shown in Figure 8. Once again, the error rate can be seen to be relatively small.

It can be seen that the quantum parity check and encoder form the upper half of our CNOT gate shown in Fig. 3. The operation of such a device would require four single photons, two of them in an entangled state. A CNOT operation can also be performed using a three-photon arrangement [8] in which a single ancilla enters the top of the diagram and exits from below, as shown in Figure 9. Although this arrangement is easier to implement, the correct results are only obtained when a single photon actually exits in each output port, which can be verified using coincidence measurements (the so-called coincidence basis). The results from the first experimental demonstration [8] of a CNOT gate for photons are shown in Figure 10. Here mode mismatch is responsible for most of the incorrect results.

The devices described above succeed with probabilities ranging from $1/4$ to $1/2$. Increasing the probability of success would require the use of larger numbers of ancilla photons [1, 2]. In addition to requiring the generation of ancilla photons in entangled states [34], the ancilla must also be detected with high efficiency. In order to avoid these difficulties, we are currently investigating the possibility of a hybrid approach [35] that combines linear optical techniques with a small amount of nonlinearity. It is expected that an approach of this kind will be able to greatly reduce the requirements for large numbers of ancilla and high detection efficiency. In particular, we have shown that the failure rate of devices of this kind can be reduced to zero using the quantum Zeno effect [35].

**IV. Single-photon source and memory**

The linear optical techniques described above are a promising method for implementing the quantum logic operations that would be required for a quantum repeater. But a source of single photons and a quantum memory would also be required for quantum repeater applications. In this section, we describe prototype experiments in which both of these devices were demonstrated.

In many respects, parametric down-conversion is an ideal way to generate single photons [9]. As illustrated in Figure 11, a pulsed laser beam incident on a nonlinear crystal will produce pairs of photons. If one member of a pair is detected, that signals the presence of the other member of the pair. A high-speed optical switch was then used to store the remaining photon in an optical storage loop until it was needed, at which time it could be switched back out of the storage loop. Although a source of this kind cannot produce photons on demand at arbitrary times, it can produce photons at specific times that can be synchronized with the clock time of a quantum computer, which is all that is required for practical applications.

Some experimental results [9] from a single-photon source of this kind are shown in Figure 12. It can be seen that the source is capable of producing and storing single photons for later use, but there was a loss of roughly 20% per cycle time in the original experiment. We are currently working on an improved version of this experiment in which the photons are stored in an optical fiber loop and special-purpose switches are used to reduce the amount of loss.

It is also possible to construct a quantum memory for photons by switching them into an optical storage loop and then switching them out again when needed [3, 4]. In this case the system must maintain the polarization state of the photons in order to preserve the value of the qubit, which is more challenging than the single-photon source described above. This can be accomplished by using a polarizing Sagnac interferometer as the switching mechanism, as illustrated in Figure 13. We have also performed a proof-of-principle experiment [4] of this kind where, once again, there were significant losses due to the optical switch.

**V. Quantum repeaters**

In the ideal case, a quantum repeater should be able to correct for all forms of errors that may occur in the transmission of a photon through an optical fiber, including phase and bit-flip errors. But as a practical matter, the dominant error source in fiber-based QKD systems is simply the loss of photons due to absorption or scattering. In the quantum key distribution systems that we have implemented, all other sources of error are negligible; there is no measurable decoherence of those photons that pass through the fiber, even when the overall absorption rate is high.

As a result, it may be sufficient to consider a quantum repeater system that compensates only for photon loss and simply ignores any other form of error. Such a system can be implemented using a simple four-qubit encoding, as shown by J. Dowling's group at the Jet Propulsion Laboratory [5]. The necessary encoding into four qubits can be done using the circuit shown in Figure 14. It can be seen that this encoding can be accomplished using a combination of CNOT logic gates and single-qubit operations, which can be easily implemented in an optical approach.

Once the qubits have been encoded in this way, the effects of photon loss can be corrected [5] using the circuit shown in Figure 15. Here a quantum non-demolition measurement is designated by the abbreviation QND, H represents a Hadamard transformation, the sigmas represent the usual Pauli spin matricies, and the polygons represent a single-photon source used to replace any photons that have been lost. QND measurements can also be implemented [29, 36] using linear optical techniques, so that the entire error correction process can be performed using the kinds of techniques that are described above.

A quantum repeater would then consist of a series of error correction circuits of this kind, separated by a sufficiently short distance $L$ of optical fiber that the probability of absorbing two or more photons in a distance L is negligibly small. Alternatively, the optical fibers could be formed into a set of loops to implement a quantum memory device as described above, where the error correction circuits would correct for the effects of photon loss and extend the storage time [3, 5].

Since the error correction circuit of Figure 15 does not correct for other types of errors, it will also be necessary to minimize the failure rate of the CNOT gates by using a sufficiently large number of ancilla photons [2] or by using a concatenated code as described by KLM [1]. It may also be possible to reduce the requirements on the number of ancilla and the detector efficiency by using a hybrid approach, such as the Zeno gates [35] that were briefly mentioned above.

## VI. Summary

In summary, we have reviewed some of the challenges faced by quantum communications systems, both past and present. Earlier difficulties associated with changes in the state of polarization and sensitivities to interferometer phase drift have been largely overcome. Although free-space systems will probably be used for special applications, their bandwidth is limited and quantum repeaters will probably be required in order to achieve the desired bandwidth and operating range. We have demonstrated several kinds of quantum logic gates [6-8], along with a prototype source of single photons [9] and a quantum memory device [4]. As shown by the group at JPL [5], these techniques can be combined with a four-qubit code to correct for the effects of photon loss and to implement a quantum repeater system. Further work will be required in order to reduce the failure rate of linear optics quantum logic gates, possibly including the development of hybrid approaches such as Zeno gates [35].

This work was supported by the Army Research Office and by IR&D funds.

**References:**

1. E. Knill, R. Laflamme, and G.J. Milburn, Nature **409**, 46 (2001).

2. J. D. Franson, M. M. Donegan, M. J. Fitch, B. C. Jacobs, and T. B. Pittman, Physical Review Letters **89**, 137901-1 to 137901-4 (2002).

3. J.D. Franson and T.B. Pittman, Proceedings of the Fundamental Problems in Quantum Theory Workshop, D. Lust and W. Schleich, Eds., Fortschritte der Physik **46**, 697-705 (1998).

4. T. B. Pittman and J. D. Franson, Physical Review A **66**, 062302-1 to 062302-4 (2002).

5. R.M. Gingrich, P. Kok, H. Lee, F. Vatan, and J.P. Dowling, Phys. Rev. Lett. **91**, 217901 (2003).

6. T. B. Pittman, B. C. Jacobs, and J. D. Franson, Physical Review A **64**, 062311-1 to 062311-9 (2001).

7. T. B. Pittman, B. C. Jacobs, and J. D. Franson, Physical Review Letters **88**, 257902-1 to 257902-4 (2002).

8. T.B. Pittman, M.J. Fitch, B.C. Jacobs, and J.D. Franson, Physical Review A **68**, 032316 (2003).

9. T. B. Pittman, B. C. Jacobs, and J. D. Franson, Physical Review A **66**, 042303-1 to 042303-7 (2002).

10. C.H. Bennett, F. Bessette, G. Brassard, L. Salvail, and J. Smolin, J. Cryptology **5**, 3 (1992).

11. J. D. Franson and H. Ilves, Applied Optics **33**, 2949-2954 (1994).

12. J. D. Franson and H. Ilves, Journal of Modern Optics **41**, 2391-2396 (1994).

13. J. D. Franson and B. C. Jacobs, Electronics Letters **31**, 232-234 (1995).

14. J. D. Franson, Phys. Rev. Lett. **62**, 2205-2208 (1989).

15. J. D. Franson, Phys. Rev. Lett. **67**, 290-293 (1991).

16. J. Rarity, private communication, 1989.

17. A.K. Ekert, J.G. Rarity, P.R. Tapster, and G.M. Palma, Phys. Rev. Lett. **69**, 1293 (1992).

18. W. Tittel, J. Brendel, H. Zbinden, and N. Gisin, Phys. Rev. Lett **84**, 4737 (2000); G. Ribordy, J. Brendel, J.-D. Gautier, N. Gisin, and H. Zbinden, Phys. Rev. A **63**, 012309 (2001).

19. C.H. Bennett, Phys. Rev. Lett. **68**, 3121 (1992).

20. P.D. Townsend, J.G. Rarity, and P.R. Tapster, Electron. Lett. **29**, 634 (1993)

21. P.D. Townsend, J.G. Rarity, and P.R. Tapster, Electron. Lett. **29**, 1291 (1993).

22. P.D. Townsend, Electronics Letters **30**, 809 (1994).

23. R. Hughes, G. Morgan, C. Peterson, J. Modern Optics **47**, 533 (2000).

24. A.T. Muller, T. Herzog, B. Huttner, W. Tittel, H. Zbinden, and N. Gisin, Appl. Phys. Lett. **70**, 793 (1997).

25. B. C. Jacobs and J. D. Franson, Optics Letters **21**, 1854-1856 (1996).

26. W.T. Buttler, R.J. Hughes, S.K. Lamoreaux, G.L. Morgan, J.E. Nordholt, and C.G. Peterson, Phys. Rev. Lett. **84**, 5652 (2000).

27. J.G. Rarity, P.R. Tapster, and P.M. Gorman, Journal of Modern Optics **48**, 1887 (2001).

28. Z. Zhao, T. Yang, Y.-A. Chen, A.-N. Zhang, and J.-W. Pan, Phys. Rev. Lett. **90**, 207901 (2003); N. Gisin, G. Ribordy, W. Tittel, H. Zbinden, Rev. Mod. Phys. **74**,

145 (2002).

29. B. C. Jacobs, T. B. Pittman, and J. D. Franson, Physical Review A **66**, 052307-1 to 052307-6 (2002).

30. T. B. Pittman, B. C. Jacobs, and J. D. Franson, Physical Review A **66**, 052305-1 to 052305-7 (2002).

31. M. Koashi, T. Yamamoto, and N. Imoto, Phys. Rev. A **63**, 030301 (2001).

32. J.W. Pan, C. Simon, C. Brukner, and A. Zeilinger, Nature **410**, 1067 (2001).

33. T.B. Pittman, B.C. Jacobs, and J.D. Franson, quant-ph/0312097; submitted to Phys. Rev. A.

34. J.D. Franson, M.M. Donegan, and B.C. Jacobs, quant-ph/0303137; submitted to Phys. Rev. A.

35. J.D. Franson, B.C. Jacobs, and T.B. Pittman, quant-ph/0401133; to be submitted to Phys. Rev. Lett.

36. P. Kok, H. Lee, and J.P. Dowling, Phys. Rev. A 66, 063814 (2002).

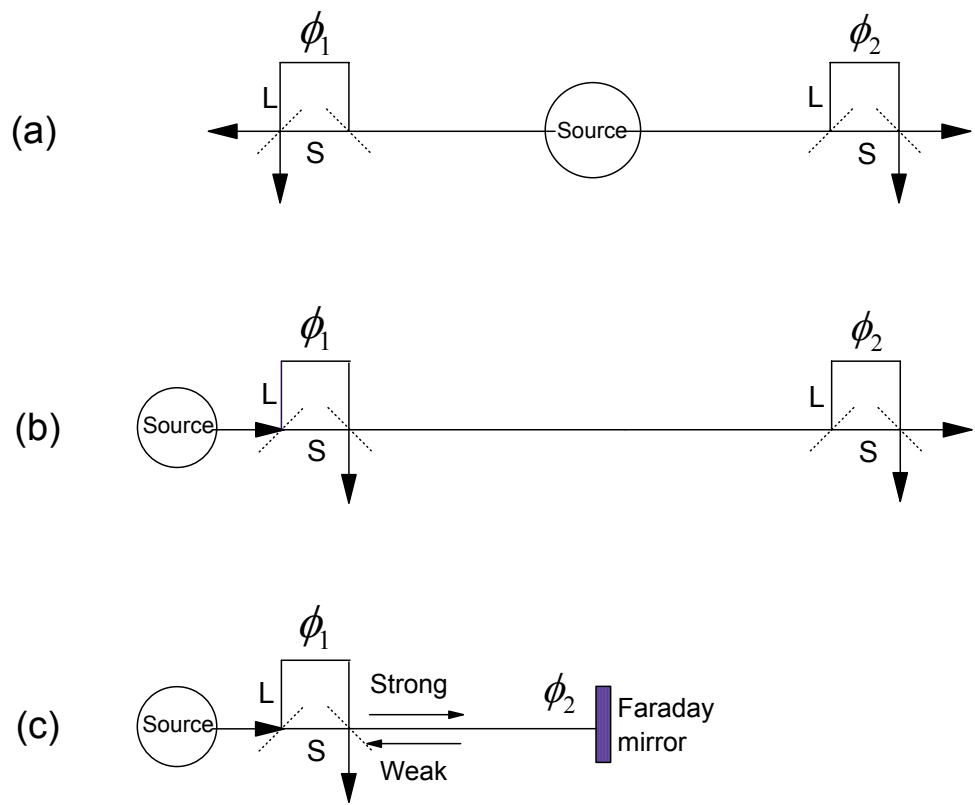

Figure 1. Evolution of interferometer-based quantum key distribution systems. (a) Nonlocal interferometer suggested by Franson in which an entangled pair of photons propagate towards two separated interferometers with a long path L and a short path S. (b) Modification by Bennett to utilize a single photon passing through two interferometers in series. (c) Plug-and-play system by Gisin's group that folds the above system in half using a Faraday mirror.

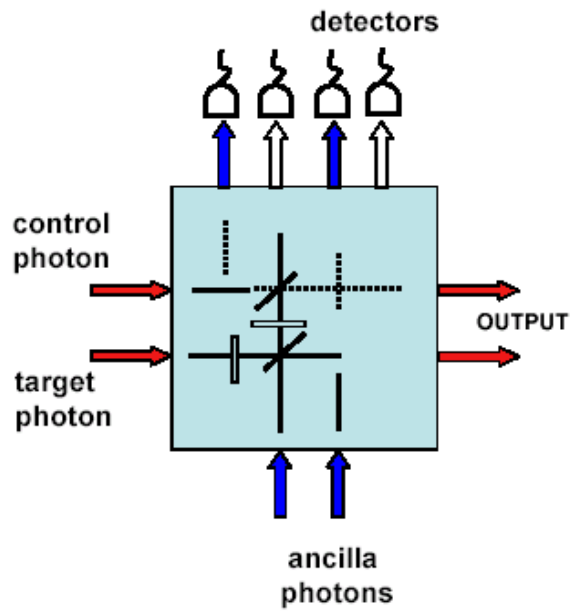

Figure 2. Basic idea behind linear optics quantum logic gates. One or more ancillla photons are mixed with two input qubits using linear elements. Post-selection based on measurements made on the ancilla will project out the correct state of the two output qubits. Feed-forward control can be used to accept additional measurement results.

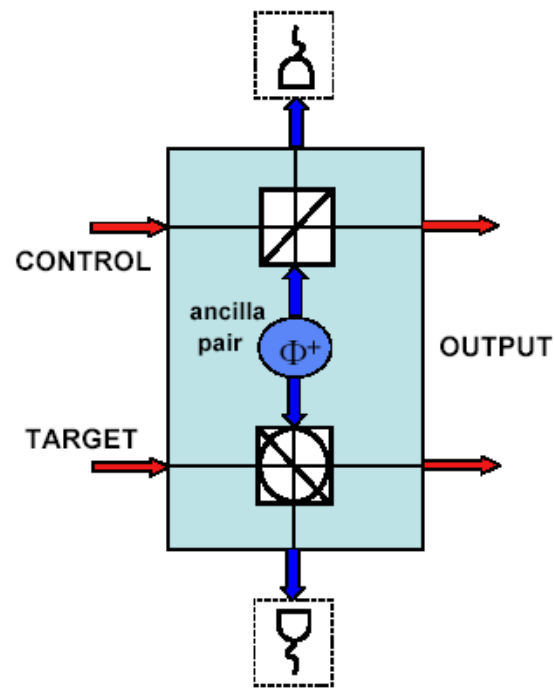

Figure 3. Controlled-NOT gate using polarization-encoded qubits. The correct logical output is obtained whenever one and only one photon is detected in both detectors, which occurs with a probability of ¼.

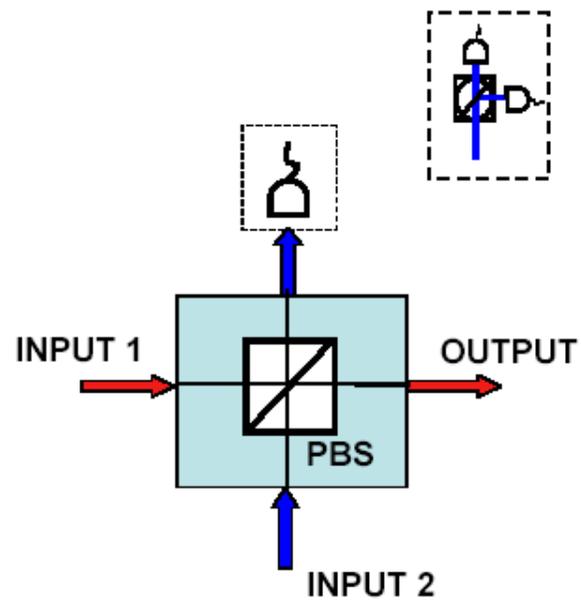

Figure 4. Implementation of a parity check operation using a polarizing beam splitter and a polarization-sensitive detector. As shown in the insert, the polarization-sensitive detector consists of a second polarizing beam splitter oriented at a 45 degree angle, followed by two ordinary single-photon detectors.

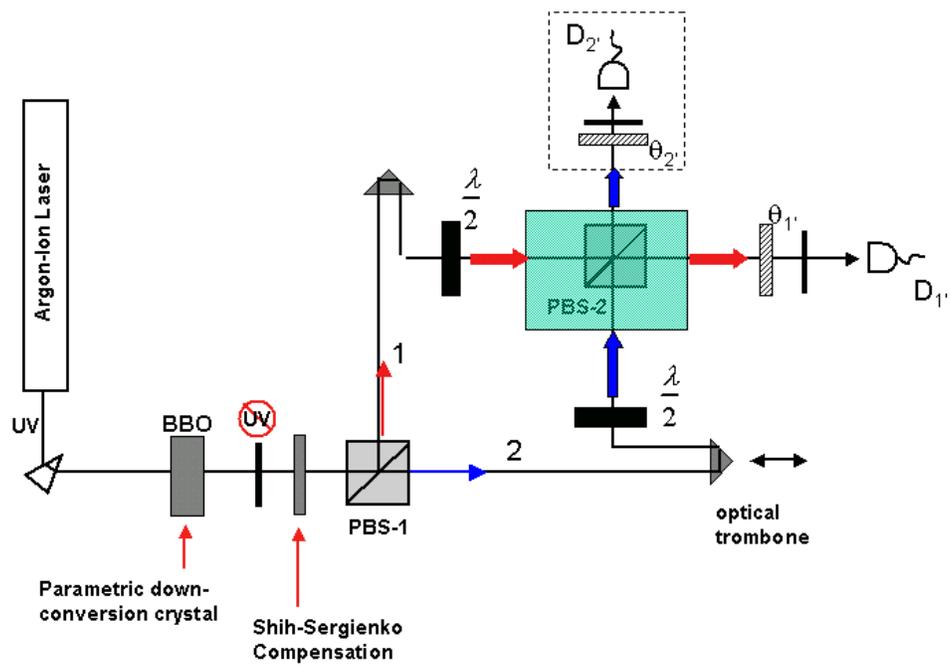

Figure 5. Experimental apparatus used to perform a demonstration of a quantum parity check and a destructive CNOT logic gate.

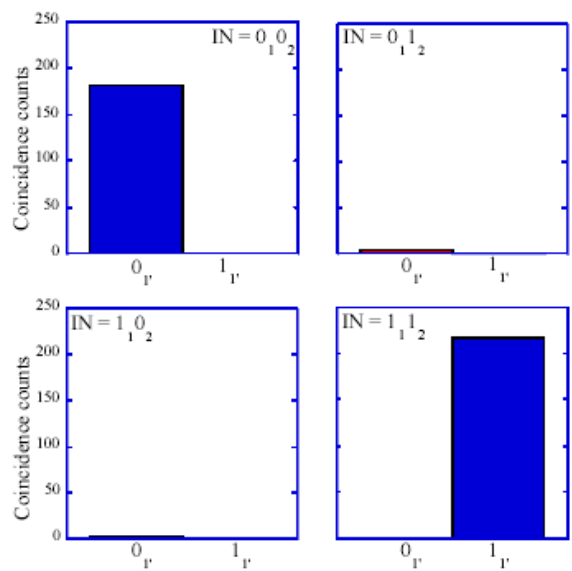

Figure 6. Experimental results from a demonstration of a quantum parity check operation.

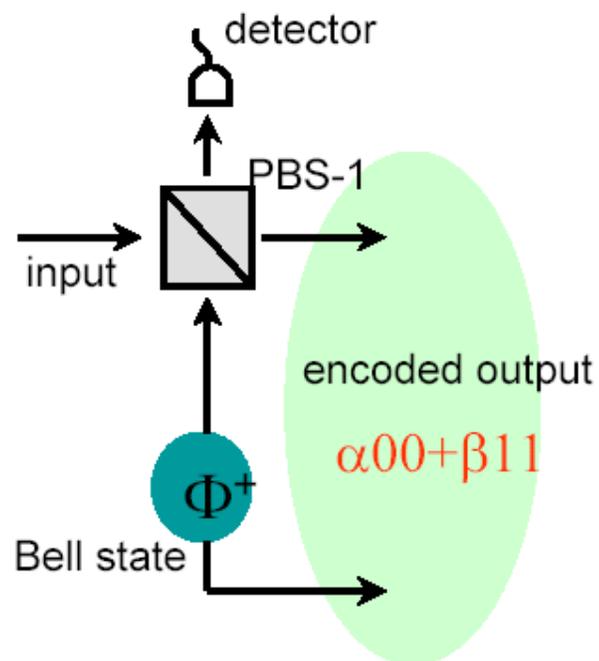

Figure 7. Implementation of a quantum encoder using a polarizing beam splitter and an entangled pair of ancilla photons.

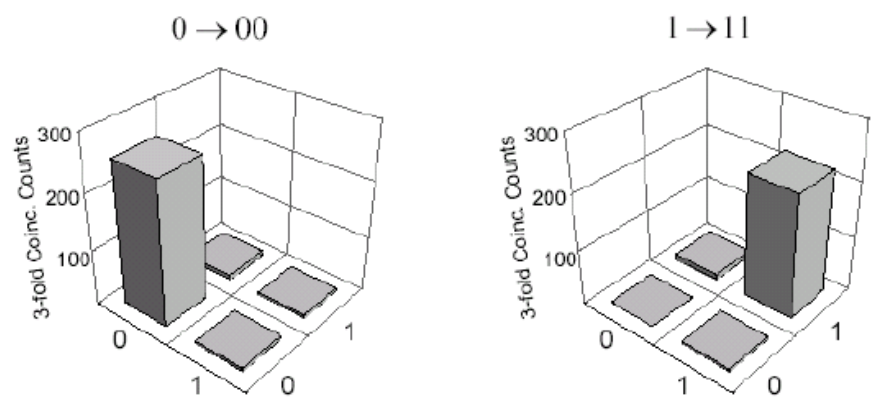

Figure 8. Experimental results from a demonstration of a quantum encoder.

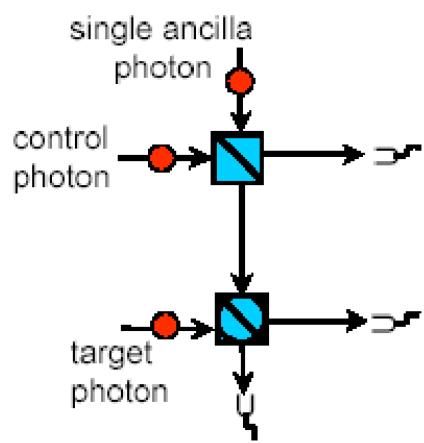

Figure 9. Implementation of a CNOT gate in the coincidence basis using a single ancilla photon.

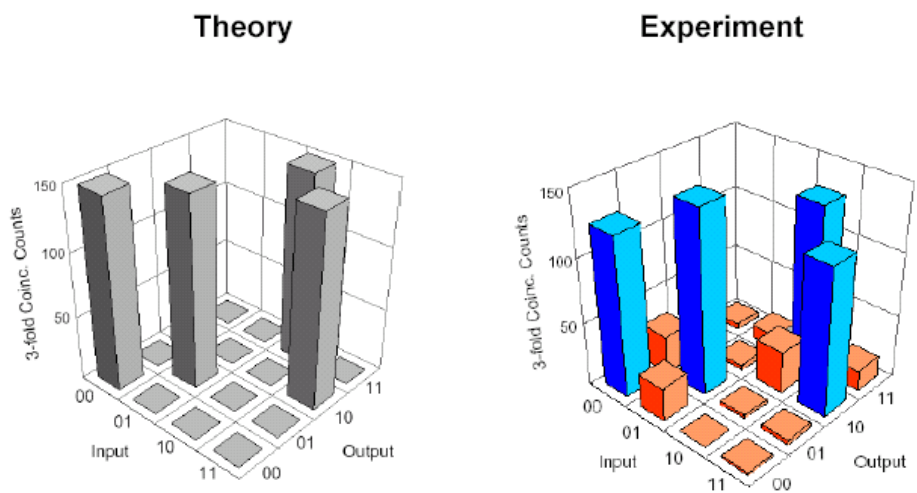

Figure 10. First experimental demonstration of a CNOT gate for single photons.

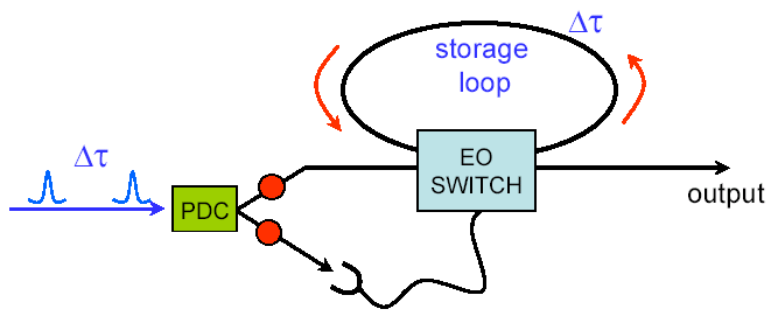

Figure 11. Implementation of a single-photon source using pulsed parametric down-conversion. The detection of one member of a pair of down-converted photons indicates the presence of the second member of the pair, which is then switched into an optical storage loop until it is needed.

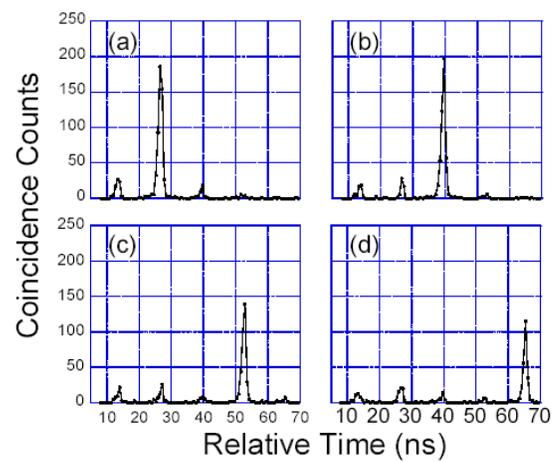

Figure 12. Experimental results from a prototype single-photon source. Figures (a) through (d) show the relative probability of switching the photon out after one through five round trips through the optical storage loop.

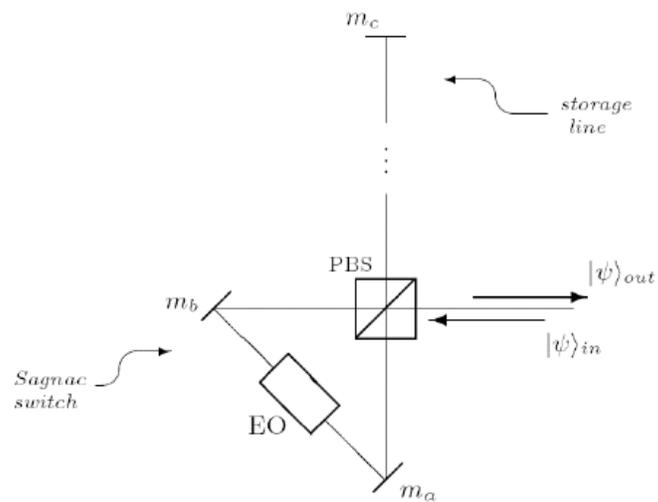

Figure 13. Polarizing sagnac interferometer used as the switching element for a single-photon memory device. A single photon can be stored in the delay line until needed and then switched out again without changing its state of polarization, aside from small technical errors.

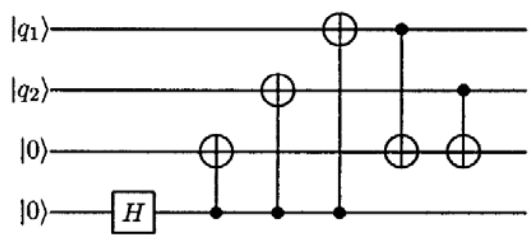

Figure 14. Circuit used to encode two logical qubits into four physical qubits, as suggested by Gingrich et al [5].

Figure 15. A circuit that can be used to correct for photon loss based on the four-qubit encoding of Figure 14, as proposed by Gingrich et. al [5].